\documentclass{openjournal}

\usepackage{lipsum}

\usepackage{xcolor}
\usepackage{textgreek}
\usepackage[utf8]{inputenc}
\usepackage[english]{babel}

\usepackage{hyperref}
\hypersetup{
    unicode, 
    colorlinks=true,
    linkcolor=linkcolor,
    citecolor=linkcolor,
    filecolor=linkcolor,
    urlcolor=linkcolor,
}
\usepackage{color,colortbl}
\definecolor{linkcolor}{rgb}{0.0,0.3,0.5}
\usepackage{tensind}
\tensordelimiter{?}
\DeclareGraphicsExtensions{.bmp,.png,.jpg,.pdf}
\usepackage{verbatim}
\usepackage[normalem]{ulem}
\usepackage{orcidlink}
\usepackage{soul}

\graphicspath{ {./figs/} }

\def\aap{Astronomy and Astrophysics}

\def\apjs{ApJS}
\def\apjl{ApJL}

\usepackage{amsmath}     

\begin{document}
\title{Faraday Depolarization Study of a Radio Galaxy Using LOFAR Two-metre Sky Survey: Data Release 2}

\author{Samantha Sneha Paul}
\email{paul.samantha.sp@gmail.com}
\affiliation{Department of Physics, Banwarilal Bhalotia College, Asansol, West Bengal, Pin: 713303, India}

\author{Abhik Ghosh\orcidlink{ https://orcid.org/0000-0002-8566-9955}}
\email{abhik.physicist@gmail.com}
\affiliation{Department of Physics, Banwarilal Bhalotia College, Asansol, West Bengal, Pin: 713303, India}

\begin{abstract}
We present a detailed depolarization analysis of the radio galaxy \texttt{ILTJ012215.21+254334.8} using polarimetric data from the \textit{LOFAR Two-metre Sky Survey} (LoTSS) Data Release 2 (DR2) catalogue. This source, with \textit{RM} $\sim$ 47 rad m$^{-2}$ and projected linear size $\sim$ 335 kpc at $z \approx 0.05$, serves as a pilot for systematic qu-fitting of mostly unresolved LoTSS sources, building on prior single-target studies that established the need for multi-component Faraday models in complex magneto-ionic media. 

Fitting five depolarization models to the LoTSS HBA (120-168 MHz) fractional polarization spectra reveals a decisively preferred three-component model (one Faraday-thin instrumental leakage, plus two external Faraday dispersions), demonstrating that LoTSS data alone can constrain moderate Faraday complexity in typical DR2 galaxies. Our results highlight turbulence and inhomogeneity in the foreground magneto-ionic medium and outline a path for population studies of LoTSS FR-I sources.
\end{abstract}

\begin{keywords}
{techniques: polarimetric, galaxies: active, galaxies: magnetic
fields, radio : galaxies, radiation mechanisms: non-thermal}
\end{keywords}

\maketitle

\section{Introduction}
\label{sec:Intro}
Radio galaxies are energetic extragalactic objects distinguished by vast radio-emitting lobes that extend well beyond the optical extent of their host galaxies. These lobes are powered by highly collimated plasma jets launched from an active galactic nucleus (AGN), itself driven by accretion onto a central supermassive black hole (SMBH) \citep{Oei24}. Their radio luminosities can reach up to \(10^{39}\,\mathrm{W/Hz}\) across frequencies from 10 MHz to 100 GHz, primarily due to synchrotron radiation emitted by relativistic electrons spiralling around magnetic field lines. Typically hosted by massive elliptical galaxies, the term \emph{radio galaxy} refers both to the host galaxy and its extended jet-lobe system. The SMBH at the core powers the jets through accretion of surrounding material \citep{Hardcastle20}. The lobes, often observed as large, symmetrical pairs, represent the most extended structures, sometimes spanning sizes of order megaparsecs \citep{Oei23}. Their radio emission is frequently highly linearly polarized \citep{begelman_blandford_rees_1984, Miley80}.  

Radio galaxies exhibit a wide range of physical sizes, from a few parsecs up to several megaparsecs, and are morphologically classified according to the Fanaroff-Riley (FR) scheme \citep{FR74}. FR~II galaxies are powerful, edge-brightened sources with luminous hotspots at the extremities of their lobes. They commonly show weak but well-collimated jets that often appear one-sided owing to relativistic beaming, with radio luminosities typically above \(\sim 10^{25}\,\mathrm{W\,Hz}^{-1}\) at 178 MHz \citep{FR74}. These systems are typically associated with quasars or more distant radio-loud galaxies. Conversely, FR~I galaxies are weaker, edge-darkened sources whose emission is brightest near the core and decreases outward. They generally lack prominent hotspots, and their jets and lobes often appear distorted. With radio luminosities below \(\sim 10^{25}\,\mathrm{W\,Hz}^{-1}\), FR~I sources are more frequently found in nearby galaxies or dense galaxy clusters environment. Although \cite{Mingo19} recently confirmed for LoTSS catalogue radio luminosity is not always a reliable way to determine if a source is edge-brightened (FR-II) or centre-brightened (FR-I). 

Extragalactic foreground radio sources have played a crucial role in advancing our understanding of the cosmic evolution of magnetic fields in the Milky Way, in external galaxies, and in the large-scale cosmic web (e.g., \cite{Taylor09}, \cite{Harvey11}, \cite{Mao10}). The study of radio polarization provides further insight into the morphology of radio galaxies and their surrounding environments \citep{Sullivan_2012, Pasetto16}. Broadband radio spectro-polarimetric observations \citep{Farnsworth_2011, Anderson15, OSullivan_2017, Pasetto18} have proven particularly effective for investigating the magneto-ionic medium in AGNs by measuring Faraday rotation measures (RMs). 

When synchrotron radiation propagates through a magneto-ionic medium, or \emph{Faraday screen}, the polarization angle experiences a rotation known as Faraday rotation (FR). Simultaneously, the polarized flux density may be reduced through processes collectively referred to as depolarization. Depolarization is generally divided into two main categories: \emph{external depolarization} and \emph{internal depolarization} \citep{Pasetto18}. Modelling these effects requires considering the structure of the magnetic field, which may range from fully ordered to a mixture of ordered and turbulent components \citep{burn66, Sokoloff98, Rossetti08, Sokollf24}.  The study of Faraday depolarization in radio galaxies provides a critical window into the magneto-ionic environments both within and surrounding these powerful extragalactic sources \citep{Strom73}.

Broadband QU-fitting of individual radio galaxies has already established that multiple Faraday components are often required to describe their polarization behaviour, and that magneto-ionic environments are typically inhomogeneous and turbulent \citep{Farnsworth_2011, Anderson15, OSullivan_2017}. Such work has largely focused either on bright, well-studied individual AGN with wide bandwidth coverage, or on small samples selected for particular environments (e.g., cluster cores, remnant sources). At higher angular resolution, resolved polarimetric imaging has revealed strong spatial variations in RM and depolarization across jets and lobes, further highlighting the complexity of the underlying magnetic field and thermal gas distributions \citep{Sullivan_2012, Pasetto18}. However, there has been relatively little effort to connect such detailed studies to the emerging large low-frequency RM grids such as LoTSS-DR2, which now provide homogeneous polarization measurements for thousands of sources \citep{Oei23}.

The specific aim here is to use LOFAR HBA data for one representative LoTSS-DR2 source to test how well standard depolarization models can be constrained from unresolved fractional Stokes spectra alone, and to characterize the multi-component Faraday structure implied by the LoTSS-only data. This single-source analysis is designed as a proof of concept for a future systematic qu-fitting study of the FR-I subsample in LoTSS-DR2, where the same methodology can be applied uniformly to a statistically meaningful population. Here, we investigate the depolarization properties of a single source drawn from the LoTSS DR2 catalogue \citep{lofar23}. Located at a redshift of 0.05, with a radio luminosity of $\sim 10^{24} \ {\rm W/Hz}$ at 144 MHz, this source can be a FR~I radio galaxy \citep{FR74}. We employ the \emph{qu-fitting} technique to model the fractional Stokes parameters as functions of wavelength squared. Recent studies have demonstrated that this approach is often more sensitive than RM synthesis for detecting multiple Faraday components \citep{Farnsworth_2011, Sullivan_2012, Sun_2015}. 

The paper is organized as follows: Section~\ref{sec:polFar} introduces the key observables of radio polarization and Faraday rotation; Section~\ref{sec:depolmodels} describes the depolarization models applied in our analysis; Section~\ref{sec:DR2source} outlines the main properties of the selected LoTSS DR2 source; Section~\ref{sec:methods} details the methodology; Section~\ref{sec:results} presents the results; and Section~\ref{sec:discuss} discusses the implications and provides prospects for future work.

\section{Polarization and Faraday Rotation}
\label{sec:polFar}
Polarization in radio galaxies offers direct evidence for synchrotron radiation, caused by relativistic electrons spiralling along magnetic fields, which naturally produces linearly polarized light. Measuring both the degree and orientation of polarization reveals the strength and structure of magnetic fields in jets and lobes, as well as information about the galaxy's orientation relative to the observer.To quantitatively describe the polarization state of electromagnetic radiation, a set of four values is utilized known as Stokes parameters: I, Q, U, and V. These parameters
are experimentally convenient as they correspond to sums or differences of measurable intensities, allowing for direct determination from observations. The complex linear polarization is defined as: 
\begin{equation}
P = Q + iU = pIe^{2i\psi} ,
\end{equation}
here, I, Q, and U are the measured Stokes parameters, and p and \(\psi\) are the fractional polarization and the polarization angle of the polarized wave, respectively, described as :
\begin{equation}
    p = \frac{S_{pol}}{I} = \sqrt{q^2 + u^2} ,
\end{equation}
Here, ${\rm S_{pol}}$ is the polarized intensity, \(q = \frac{Q}{I}\) and \(u = \frac{U}{I}\) as the fractional values of the Stokes parameters respectively, and \(\psi = \frac{1}{2} \arctan\frac{u}{q}\).

The study of the polarization states of radio emissions enables us to analyse two important effects: Faraday rotation and Faraday depolarization \citep{Anderson15,Pirasthesis2024}. Faraday rotation is caused by the rotation of the plane of polarization of electromagnetic waves as they journey across vast cosmic distances interacting with magnetized medium. Crucially, the extent of
this rotation directly encodes vital information about the magneto-ionic medium—the ionized gas and embedded magnetic fields. This makes Faraday rotation an exceptionally powerful tool for probing magnetic fields not just within galaxies, but also in the vast and tenuous intergalactic medium, providing unique insights into the large-scale magnetization of the Universe that cannot be obtained
through other means \citep{Brentjens05}. To accurately characterize the Faraday rotation for these distinct regions of polarized emission, the concept of Faraday depth \((\varphi)\) is utilized.  Faraday depth is defined as

\[
\varphi(s) = 0.81 \int_0^s n_e(s') \, B_\parallel(s') \, ds' \,\, \left[\text{rad} \, \text{m}^{-2}\right],
\]

where $s$ is the physical path length coordinate measured from the observer ($s = 0$) towards the source, $n_e$ is the thermal electron density in $\text{cm}^{-3}$, and $B_\parallel$ is the component of the magnetic field in $\mu\text{G}$ parallel to the line of sight, taken to be positive when pointing towards the observer \citep{Ferri21}.

Within an extended source, emission at different physical depths contributes polarized radiation with distinct \(\varphi\), enabling multi-component Faraday spectra even without beam depolarization.

For a simple foreground screen (all emission at \(\varphi > \) screen), the observed rotation measure \(\mathrm{RM}\) equals the screen Faraday depth. In general, \(\mathrm{RM}\) is the first moment of the Faraday dispersion function \(\mathcal{F}(\varphi)\), while \(\varphi\) traces the full depth structure. We use both quantities interchangeably when referring to a dominant single-screen case.
In such an ideal case, the observed polarization angle \(\psi\) is directly determined by the source's intrinsic polarization angle \(\psi_0\) and the Faraday rotation measure (RM), following the simple linear relationship \citep{Anderson16}:
\begin{equation}
    \psi = \psi_0 + RM \lambda^2
\end{equation}
Here, ${\rm{\lambda^2}}$ represents the square of the observing wavelength, illustrating how the rotation of the polarization plane is directly proportional to both the RM of the intervening medium and the square of the wavelength at which the observation is made \citep{burn66, Brentjens05}.

In real astrophysical cases, the magnetic field is not perfectly uniform, leading to a significant reduction in the polarization percentage with increasing wavelength. This effect is known as Faraday depolarization \citep{Sullivan_2012, OSullivan_2017, Pasetto18, Passeto21, Pirasthesis2024}.

\section{Depolarization}
\label{sec:depolmodels}
Depolarization occurs because variations in the magnetic field or electron density along the line of sight, or across the telescope beam, cause the polarization angle to rotate by different amounts in different regions, reducing the net observed polarization \citep{burn66, Goodlet04}. When these diverse polarization orientations are averaged across the observing beam, they tend to cancel each other out, diminishing the overall polarization signal. Hence, two types of depolarization come into play: internal and external \citep{Pasetto18}. The remaining polarization percentage is effectively utilized to gauge the degree of order and orientation within the magnetic field, whether it resides in the radio source itself or in the intervening inter galactic medium (IGM). The complex polarized signal observed from radio sources is fundamentally described by a model that accounts only for Faraday rotation and does not depolarize the signal is given as \citep{burn66} :
\begin{equation}
    p(\lambda^2;p_0;\psi_0;RM) = p_0 e^{2i(\psi_0+RM\lambda^2)} 
    \label{eq:burn}
\end{equation}
Here, \(p_0\) represents the intrinsic fractional polarization of the radiation, and \(\psi_0\) is the intrinsic polarization angle at the source of the emission. The term \(RM\lambda^2\) describes the Faraday rotation caused by the foreground magneto-ionic medium, where RM is the rotation measure and \(\lambda^2\) is the square of the observing wavelength. This model is also referred as `Faraday thin' \citep{burn66}. In real scenarios, the magnetic field is not perfectly ordered, there exists random and uniform fields causing depolarization \citep{Goodlet04}. The various depolarization contributions are thus added to this fundamental equation \ref{eq:burn}.

When the polarized radiation passes through an external magneto-ionic medium with a turbulent magnetic field, the depolarization caused is called `external Faraday dispersion'. This is modelled as external depolarization component and is represented by the following equation :
    \begin{equation}
    p(\lambda^2;p_0;\psi_0;RM;\sigma_{RM}) = p_0 e^{2i(\psi_0+RM\lambda^2)} e^{-2\sigma^2_{RM}\lambda^4}
    \end{equation}
Here, \(\sigma_{RM}\) is the Faraday dispersion about the mean RM across the radio source on the sky.

In a specific scenario, where the region producing the synchrotron radiation and the area causing the Faraday rotation are co-spatial, the resulting depolarization is internal, referred to as `differential Faraday rotation' \citep{Sokoloff98}. However, in this particular model, the magnetic field is assumed to be purely regular, meaning any turbulent magnetic field component is neglected. Under these conditions, radiation from different depths within the source will undergo varying amounts of Faraday rotation, as the path length through the magnetized medium differs for each region. This is modelled as a `Faraday thick' component. It is represented by the equation :
\begin{equation}
    p(\lambda^2;p_0;\psi_0;R) = p_0 e^{2i(\psi_0+\frac{1}{2}R\lambda^2)} \frac{\sin R\lambda^2}{R\lambda^2}
\end{equation}
Here, the R is the Faraday Depth along the region and the observable RM is equal to \(\frac{1}{2}R\) in this case. Furthermore, depolarization can also arise when there is a gradient in the Faraday depth \citep{OSullivan_2017} across the source itself, or within a foreground screen that is local to the emitting region, even if the magnetic field within that region is considered uniform. In such instances, the Faraday depth effectively takes the form of a Rotation Measure (RM) that varies across the telescope's observing beam. This spatial variation in RM causes different parts of the signal within the beam to have distinct polarization orientations, which then tend to cancel each other out when averaged, resulting in observed depolarization.
In general, depolarization towards longer wavelengths can result from a combination of factors, including the mixing of the emitting and rotating media, as well as the limited spatial resolution of our observations (for more details see \citet{Sokoloff98}). 

Note, this implementation of the Faraday-thick (Burn-slab) model omits a foreground Faraday rotation term (\textit{RM\_fg}), following \citet{Sokoloff98} and as implemented in the version of RM-Tools used here. For instrumental leakage near RM $\approx$ 0, this approximation is reasonable, as ionospheric RM shifts are small (\(< 3\) rad m$^{-2}$). However, for astrophysical components, this may bias the recovered Faraday depth if a significant foreground screen is present; we discuss this further in Section \ref{sec:results}.

Moreover, for an external Faraday dispersion component, the fractional polarization scales as, \[
|p(\lambda^2)| = p_0 \exp\left(-2 \sigma_{\text{RM}}^2 \lambda^4\right), \] \citep{burn66}, so that even intrinsically 100\% polarized emission is exponentially suppressed once \( 2 \sigma_{\text{RM}}^2 \lambda^4 \gg 1 \). For the LoTSS HBA band (120–168 MHz, $\lambda^2 \approx 3.2 \, \rm{m}^2 \, - \, 6.3 \, \rm{m}^2$), this means that components with $\sigma_{\rm RM} \gtrsim 0.1 \, - \, 0.2 \,\, \text{rad} \, \text{m}^{-2}$ are already heavily depolarized ($|p|/p_0 \lesssim 10^{-2}$) across most of the band, and for $\sigma_{\rm RM} \gtrsim 0.3 \, - \, 0.5 \,\, \text{rad} \, \text{m}^{-2}$ they would be effectively undetectable at LoTSS sensitivity ($|p|/p_0 \ll 10^{-3}$) even if intrinsically strongly polarized.

\section{The LoTSS DR2 Catalogue and the Selected Source}
\label{sec:DR2source}
The LOFAR Two-metre Sky Survey (LoTSS) Data Release 2 (DR2) has compiled a comprehensive catalogue of 2,461 extragalactic sources with high-precision Rotation Measure (RM) values, covering 5,720 square degrees of the sky \citep{lofar23}. This extensive dataset, with a polarized source density of approximately 0.43 sources per square degree, was constructed by deriving linear polarization and RM properties using RM synthesis from Stokes Q and U channel images. Observations were conducted at an angular resolution of 20 arcseconds, spanning a frequency range of 120 to 168 MHz with a channel bandwidth of 97.6 kHz. The study found that roughly \(0.2\%\) of total intensity sources (brighter than 1 mJy beam\(^{-1}\)) exhibited polarization, with a median detection threshold of 0.6 mJy beam\(^{-1}\) and a median RM uncertainty of 0.06 rad m\(^{-2}\) (though a systematic uncertainty of up to 0.3 rad m\(^{-2}\) remains possible after ionospheric correction). The detected sources showed a median degree of polarization of \(1.8\%\), ranging from \(0.05\%\) to \(31\%\) \citep{lofar23}. The LoTSS DR2 RM Grid Catalogue primarily contains Fanaroff-Riley Class II (FR-II) objects, with fewer than 10\% classified as Fanaroff-Riley Class I (FR-I) sources. With a luminosity cut-off of \(< 10^{25.5}\) W Hz\(^{-1}\), there are approximately 222 sources in the LoTSS-DR2 catalogue, most of which are expected to be FR-Is \citep{FR74}. However, recent study shows \citep{Mingo19}, there can be a large population of low-luminosity FR-IIs below the expected FR break. Hence, the relative population of FR-I and FR-IIs may differ.

The present work is explicitly designed as a pilot for a larger qu-fitting campaign on the LoTSS-DR2 FR-I like population. For this purpose, a single source is preferred to satisfy three criteria: (i) it should be representative of the catalogue in terms of luminosity and size, (ii) it must have a robust polarization and RM measurement in DR2, and (iii) its angular and projected linear size should be large enough that depolarization effects due to propagation through an extended magneto-ionic medium are expected to be significant.

\begin{figure}[htbp]
    \centering
    \includegraphics[width=0.49\textwidth]{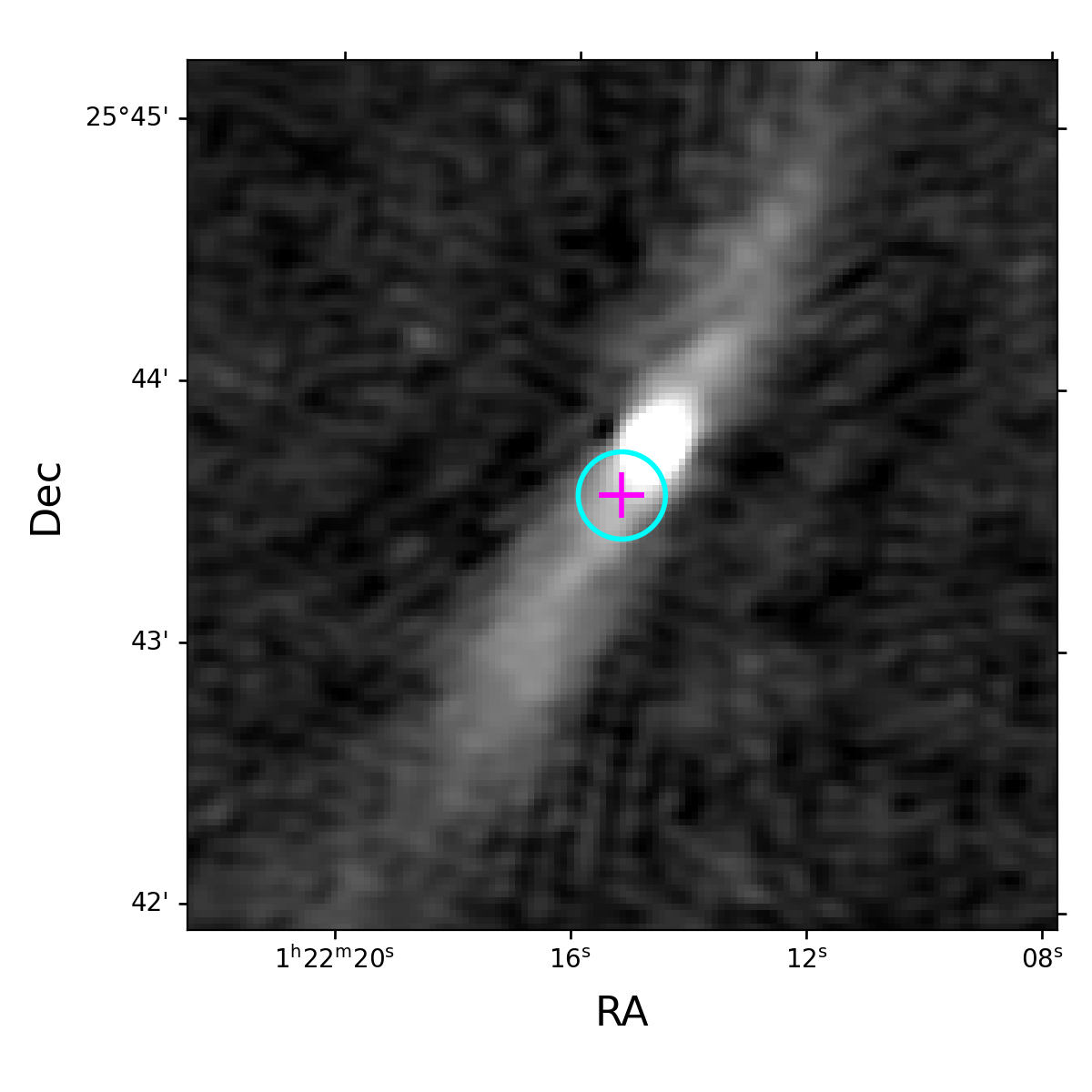}
    \includegraphics[width=0.49\textwidth]{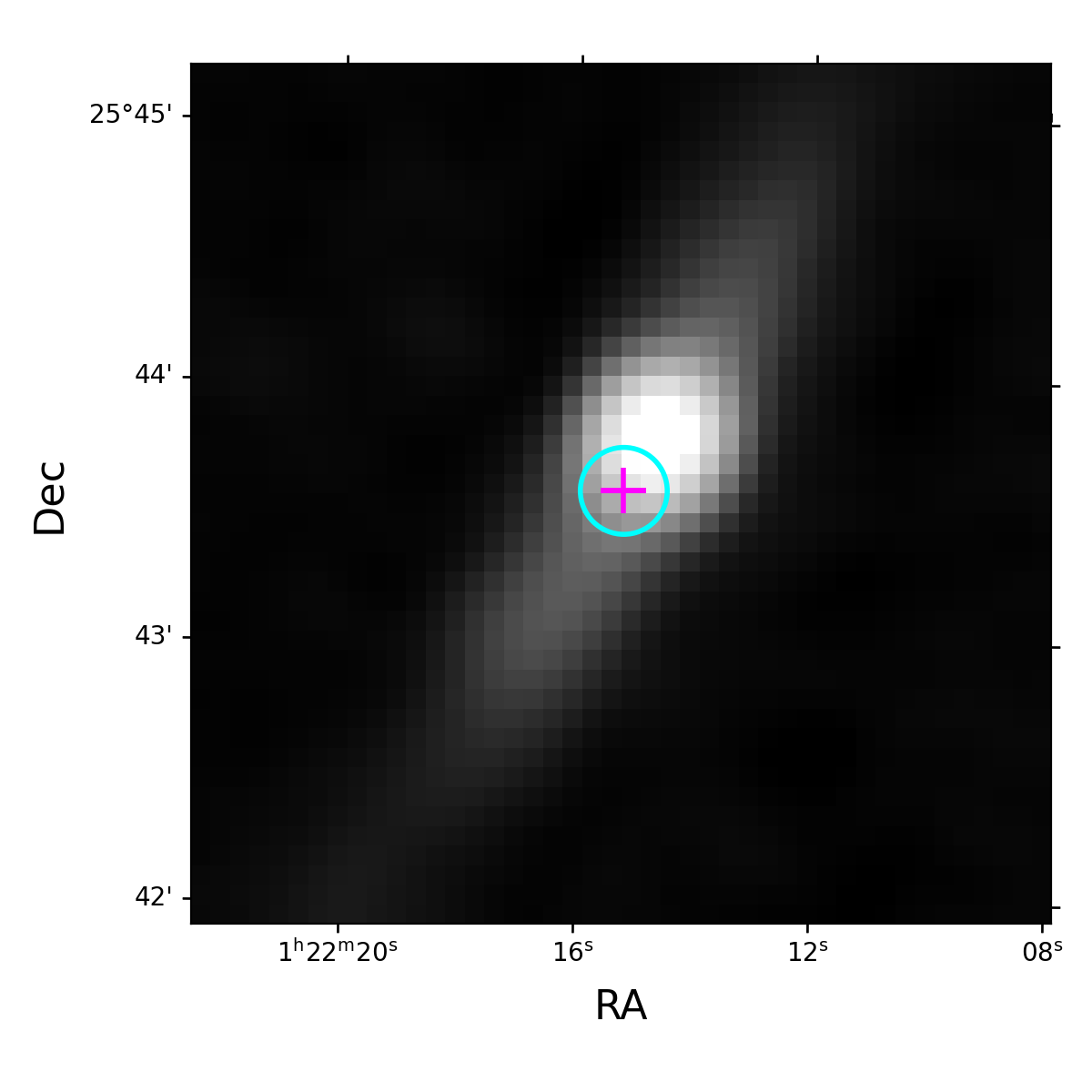}
    \caption{LOFAR image of the source ILTJ012215.21-254334.8 at resolution of $6^{''}$ (left panel) and $20^{''}$ (right panel). The cyan circle represents the $20^{''}$ resolution beam at the location of the polarized detection (marked with a magenta `+' sign). The LoTSS DR2 RM Grid catalogue was derived from the LoTSS DR2 Stokes Q and U frequency cubes at an angular resolution of $20^{''}$ across a frequency range of 120 to 168 MHz \citep{Shimwell22}.}
    \label{fig:img_LoTSS20}
\end{figure}

In this pilot study, we therefore focus on ILTJ012215.21-254334.8, which satisfies these criteria. It has an estimated luminosity of \(1.65 \times 10^{24}\) W Hz\(^{-1}\) at 144 MHz, RM of \(-47.15 \pm 0.05\) rad m\(^{-2}\), and a projected linear size of 335 kpc, i.e., very close to the LoTSS-DR2 median linear size of $\sim$ 300 kpc \citep{lofar23}. 
The source has a total angular size of $\sim 25\arcsec$ in Stokes I, comparable but not identical to the $20\arcsec$ LoTSS beam. The LoTSS–DR2 RM Grid spectrum is extracted from the single pixel corresponding to the peak polarized flux density (20\arcsec\ resolution), which is offset from the Stokes I core, and therefore represents the polarization properties of a single, core-dominated line of sight rather than the entire jet–lobe system (Figure \ref{fig:img_LoTSS20}). With an estimated luminosity of \(1.65 \times 10^{24} \, \mathrm{W \, Hz^{-1}}\) at 144 MHz (below the classical FR break of \(10^{25} \, \mathrm{W \, Hz^{-1}}\)), this is classified as an FR-I radio galaxy \citep{FR74}. The polarization fraction of approximately $3 \%$ and RM are typical of the catalogued sources, so it can be regarded as representative of the dominant population of low-luminosity FR-I-like radio galaxies in LoTSS-DR2. By analysing this object in detail, the intention is to (a) validate the qu-fitting setup, priors, and instrumental-leakage treatment under realistic LoTSS conditions, and (b) identify which depolarization parameters and model classes are actually constrained by the data for a typical source. These lessons will directly inform the design and interpretation of a forthcoming study of the full DR2 FR-I sample.

A few key parameters of the selected source are listed in Table~\ref{tab:source_simple}.

\begin{table}[ht]
    \centering
    \caption{\textbf{LoTSS DR2 Source: ILTJ012215.21+254334.8}}
    \label{tab:source_simple}
    \begin{tabular}{l c}
    \hline
    \hline
    Rotation Measure (RM) & \(-47.15 \pm 0.05\) rad m\(^{-2}\) \\
    Polarized Intensity (P) & \(4.6 \pm 0.13\) mJy beam\(^{-1}\) \\
    Fractional Polarization (\(p_0\)) & 2.99 \% \\
    Total Intensity (I) & \(154.5 \pm 0.18\) mJy beam\(^{-1}\) \\
    Radio Luminosity at 144 MHz (\(L_{144}\)) & \(1.65 \times 10^{24}\) W Hz\(^{-1}\) \\
    Linear Size ($\ell$) & 334.94 kpc \\
    Angular Size & 25" \\
    Redshift (z) & 0.0544 \\
    Morphology & FR-I \\
    \hline
    \end{tabular}
\end{table}

Here, RM is the Rotation Measure; P is the linear polarization; \(p_0\) is the fractional polarization; I is the total intensity; \(L_{144}\) is the luminosity at 144 MHz; $\ell$ is the projected linear size; and z is the redshift. The total projected linear size of $\sim 335 \, \text{kpc}$ refers to the full jet–lobe structure seen in Stokes I. In this work, however, we model only the polarization from the single $20\arcsec$ beam at the polarized peak, i.e., from the core/inner-jet region, and not from the extended, low-surface-brightness jets.

The fractional $q(\lambda^{2})$ and $u(\lambda^{2})$ spectra were extracted from the LoTSS--DR2 RM Grid catalogue at the single pixel corresponding to the peak polarized flux density position of the source (20\arcsec\ resolution). This represents the polarization properties of the mostly unresolved source as a whole, averaged over the beam. Figure \ref{fig:img_LoTSS20} shows the LOFAR image of the analysed source at resolution of $6^{''}$ and $20^{''}$. The cyan circle represents the $20^{''}$ resolution beam at the location of the polarized detection.

\section{Methods}
\label{sec:methods}
\subsection{Depolarization models}
When analysing LOFAR polarization measurements, it is essential to account for instrumental polarization, which typically appears as a component centered near a Faraday depth of $0 \, \mathrm{rad \ m^{-2}}$ in the Faraday spectrum. Due to ionospheric variations, this instrumental polarization can spread across a range of Rotation Measure (RM) values \citep{Jelic14, Mevius18}. Therefore, this effect must be carefully modelled; otherwise, it may contaminate the detection of genuine polarized emission.  

To accurately fit the observed Stokes $Q$ and $U$ values for the selected source, we considered both single and two-component models, namely: Faraday-thin (T), Faraday-thick (Tk), and external Faraday dispersion (ED), as well as combinations of these. The three model types are described as follows:

\begin{equation}
    p_T(\lambda^2;p_0;\psi_0;RM) = p_0 e^{2i(\psi_0+RM\lambda^2)} 
    \label{eq:thin}
\end{equation}
\begin{equation}
    p_{ED}(\lambda^2;p_0;\psi_0;RM;\sigma_{RM}) = p_0 e^{2i(\psi_0+RM\lambda^2)} e^{-2\sigma^2_{RM}\lambda^4}
    \label{eq:ED}
\end{equation}
\begin{equation}
    p_{Tk}(\lambda^2;p_0;\psi_0;R) = p_0 e^{2i(\psi_0+\frac{1}{2}R\lambda^2)} \frac{\sin R\lambda^2}{R\lambda^2}
    \label{eq:thick}
\end{equation}
\par
where, \(\lambda^2\) is the variable and \(p_0, \psi_0, RM, \sigma_{RM}, RM= \frac{1}{2}R\) are the parameters. These models are combined in the five models that are listed in the table \ref{table:depolmodel} given below \citep{Piras24}. We use them to find out the best fit depolarization model that fits the observed q, u spectra for the source. 

\begin{table}[ht]
    \centering
    \caption{\textbf{A summary of the models used in qu fitting}}
    \begin{tabular}{|| c c c ||}
    \hline
    Model & Description & Equations \\
    \hline\hline
    1ED & 1 External Faraday Dispersion & \ref{eq:ED} \\
    1T+1ED & Faraday Thin + 1 External Faraday Dispersion & \ref{eq:thin} + \ref{eq:ED} \\
    1T+2ED & Faraday Thin + 2 External Faraday Dispersion & \ref{eq:thin} + \ref{eq:ED} + \ref{eq:ED} \\
    1Tk+1ED & Faraday Thick + 1 External Faraday Dispersion & \ref{eq:thick} + \ref{eq:ED} \\
    1Tk+2ED & Faraday Thick + 2 External Faraday Dispersion & \ref{eq:thick} + \ref{eq:ED} + \ref{eq:ED} \\
    \hline
    \end{tabular}
    \label{table:depolmodel}
\end{table}

\subsection{The priors}
In qu-fitting, setting good starting assumptions (priors) for the model parameters is important because they help guide the fitting process in the right direction. The selection of these priors is often based on the observed Faraday spectrum of the source \citep{Pirasthesis2024}. For example, when combining multiple models, The total initial polarization fraction $\sum p_0$ across all astrophysical components must satisfy $\sum p_0 \leq 1$. The instrumental leakage term is excluded from this physical constraint, and each initial polarization angle \(\psi_0\) is limited between 0 and 180 degrees for all the models considered in Table \ref{table:depolmodel}. 

Specific constraints are applied based on the nature of the depolarization models:

Instrumental polarization, which often appears near Faraday depth 0 ${\rm rad \ m^{-2}}$ in LOFAR polarization observations, also requires specific prior constraints. Instrumental polarization can manifest across a range of RM values because it arises from RM fluctuations caused by the ionosphere throughout the observation \citep{Mevius18}. The instrumental polarization leakage is modelled as a Faraday-thin or top-hat Faraday thick (Burn-slab) component with its Rotation Measure confined near zero ($-3 \, {\rm rad \ m^{-2}}$ to $1.5 \, {\rm rad \ m^{-2}}$ for Faraday thin and $-6 \, {\rm rad \ m^{-2}}$ to $3.0 \, {\rm rad \ m^{-2}}$ for thick components respectively), capturing the typical RM range where leakage occurs in LOFAR data \citep{lofar23}. We adopt a uniform prior to constraint the instrumental polarization. This method allows for the separation of intrinsic source polarization from instrumental artifacts, enhancing the accuracy and reliability of the resulting Faraday rotation measures and polarization parameters.

For models involving a single external Faraday dispersion (1ED), the Rotation Measure (RM) is allowed to vary within the range $\pm 30\%$ of the LoTSS catalogue RM values \citep{lofar23}. Here, we also used an uniform prior.

If the model includes two external Faraday dispersion components (2ED), the uniform prior range is chosen such that one RM component is constrained within $\pm 30\%$ of the LoTSS catalogue RM, while the other is allowed a broader range, for example, from $-150 \, {\rm rad \ m^{-2}}$ to $150 \, {\rm rad \ m^{-2}}$. Additionally, the width of the Rotation Measure, $\sigma_{\rm RM}$, is forced to be positive.

\subsection{qu-fitting}
To analyse the observed Stokes $Q$ and $U$ values, along with their uncertainty, a one-dimensional qu-fitting technique was applied to the LoTSS DR2 polarization spectra of the source\footnote{https://lofar-mksp.org/data/}. The LoTSS DR2 RM Grid lacks reliable channelized Stokes I spectra\footnote{\url{https://github.com/sposullivan/LoTSS-RM-Grid/blob/main/example_RMtable_PolSpectra.ipynb}}, so we construct a model $I(\nu)$ from the catalogue 144 MHz flux density assuming a spectral index $\alpha=-0.7$. The Stokes I uncertainty $\sigma_I$ is taken as the catalogue measurement uncertainty, held constant across all frequency channels.

The fractional uncertainties are then computed as:
$$
\sigma_q \approx |q| \sqrt{\left( \frac{\sigma_Q}{Q} \right)^2 + \left( \frac{\sigma_I}{I} \right)^2}, \quad
\sigma_u \approx |u| \sqrt{\left( \frac{\sigma_U}{U} \right)^2 + \left( \frac{\sigma_I}{I} \right)^2}.
$$

This conservative approach ensures that $\chi^2_\mathrm{red}$ and Bayes factor model selection are driven by the spectral shape of $q(\lambda^2)$ and $u(\lambda^2)$, making the analysis robust to absolute polarization scale uncertainties while remaining sensitive to depolarization signatures.

We used the RM-TOOLS software\footnote{https://github.com/CIRADA-Tools/RM-Tools/tree/master} \citep{Purcell20}, which incorporates the MultiNest algorithm \citep{Feroz_2008, Feroz_2009, Feroz_2019}. MultiNest, implemented via PYMULTINEST \citep{Buch14}, is employed to efficiently explore the parameter space of the chosen model, identify the best-fit parameters, and estimate their associated uncertainties. The algorithm provides the posterior distribution for the model parameters and the Bayesian evidence ($Z$) for the model, which quantifies the probability of observing the data given a specific model. 

To compare and select between different models (e.g., model “a” versus model “b”), the Bayes odds ratio $(\Delta \ln Z)$ is used as a metric for model selection. The Bayes odds ratio is defined as:

\begin{equation}
    \Delta \ln Z = \ln Z_a - \ln Z_b = \ln \frac{Z_a}{Z_b}
\end{equation}

Here, $Z_a$ and $Z_b$ are the Bayesian evidences of models a and b, respectively. To evaluate the performance of different models in qu-fitting, we follow established statistical guidelines, such as those proposed by \citet{Kass95}, which rely on the Bayes odds ratio $(\Delta \ln Z)$. By computing the odds ratio between the model exhibiting the highest Bayesian evidence and all other candidate models, the model with the lowest resulting Bayesian odds ratio value is identified as the best-fitting representation of the observed data. 

In addition to the Bayesian odds ratio, the qu-fitting software also calculates the reduced chi-squared $(\chi^2_{\text{red}})$ for each model. This metric assesses the goodness of fit, taking into account the model's complexity. It is defined as:

\begin{equation}
\chi^2_{\text{red}} = \frac{1}{\text{DoF}} \sum_{i=1}^{N} \left[ 
\left( \frac{q_i - q_{\text{model}, i}}{\sigma_{q_i}} \right)^2 + 
\left( \frac{u_i - u_{\text{model}, i}}{\sigma_{u_i}} \right)^2 
\right]
\end{equation}

Here, DoF represents the number of degrees of freedom, $N$ is the number of data points (equivalent to the effective number of frequency channels), $q_{\text{model}}$ and $u_{\text{model}}$ are the values predicted by the model, and $\sigma_q$ and $\sigma_u$ are the uncertainties associated with q, u respectively \citep{Piras24}.

Model comparison follows the \citet{Kass95} scale for \( 2 \Delta \ln Z \), where values \( < 2 \) are barely worth mentioning, values between \( 2 \) and \( 6 \) indicate positive evidence, values between \( 6 \) and \( 10 \) indicate strong evidence, and values \( > 10 \) indicate very strong evidence against the alternative. Thus, \(2 \Delta \ln Z > 10 \) decisively favours the reference model. Here, \( 2 \Delta \ln Z > 10 \) indicates very strong evidence for \( 1T+2ED \) over alternatives \citep{Kass95}).

\section{Results}
\label{sec:results}
The comparison of different models based on the odds ratio and the reduced chi-squared statistic (\(\chi^2_\mathrm{red}\)) indicates that the best description of the depolarization behaviour of the source is provided by the \( \text{1T+2ED} \) model. This particular model includes three polarization components: one Faraday-thin term corresponding to instrumental leakage, located at a rotation measure (RM) of about $-0.3 \, {\rm rad \ m^{-2}}$, and two external depolarization (2ED) components, centred near \(\mathrm{RM} \approx -47 ~\mathrm{rad \ m^{-2}}\), which likely represent the astrophysical polarized emission \citep{lofar23}. The astrophysical emission is better represented by the superposition of two external depolarization components at similar RMs, each capturing slightly different aspects of turbulence or inhomogeneities in the magneto-ionic medium along the line of sight \citep{Sokoloff98, OSullivan_2017}. 

The fit is supported by both the Bayesian odds ratio, which quantifies the relative probability of competing models, and the \(\chi^2_\mathrm{red}\), which indicates the goodness-of-fit compared to observational uncertainties (Table \ref{table:odd-ratio-ranking}).

\begin{table}[ht]
    \centering
    \caption{\textbf{Summary of best-fitting metrics for the qu-fitting models. According to \citet{Kass95}, \( 2 \Delta \ln Z > 10 \) provides very strong evidence in favor of \( 1T+2ED \) compared to other models.}
    Column (1): model name,
    Column (2): number of free parameters ($N_{\text{free}}$),
    Column (3): number of components ($N_{\text{comp}}$),
    Column (4): degrees of freedom (DoF),
    Column (5): reduced chi-squared ($\chi^2_{\text{red}}$),
    Column (6): logarithm of the Bayesian evidence (${\rm ln(Z)}$), 
    Column (7): odds ratio expressed as $2\Delta {\rm ln(Z)}$),
    Rows are sorted by increasing odds ratio}
    \begin{tabular}{|| c c c c c c c ||}
    \hline
    Model & N$_{\text{free}}$ & N$_{\text{comp}}$ & DoF & $\chi^2_{\text{red}}$ & ln(Z) & 2$\Delta$ln(Z) \\
    \hline\hline
    1T+2ED   & 11 & 3 & 814 & 2.12 & 1385.46 & 0 \\
    1Tk+2ED  & 11 & 3 & 814 & 2.12 & 1377.38 & 16.16 \\
    1T+1ED   & 7  & 2 & 818 & 2.77 & 1154.35 & 462.23 \\
    1Tk+1ED  & 7  & 2 & 818 & 2.77 & 1148.94 & 473.05 \\
    1ED      & 4  & 1 & 821 & 2.94 & 1097.37 & 576.18 \\
    \hline
    \end{tabular}
    \label{table:odd-ratio-ranking}
\end{table}


Note here the Faraday-thick (Burn-slab) model omits a foreground \textit{RM$_{\rm fg}$} term 
\citep{Sokoloff98}, where effective RM$_{\rm Tk} = R/2$ is tied to dispersion origin. 
This limits flexibility for ionospheric-shifted instrumental leakage. The modest degeneracy 
1T+2ED ($\Delta\ln Z = 0$) vs 1Tk+2ED ($\Delta\ln Z = 16$) partly reflects this limitation.

To test this concern directly, we re-implemented \textit{RM$_{\rm fg}$} in thick models 
(1Tk+1ED, 1Tk+2ED) with RM$_{\rm Tk} = {\rm RM}_{fg} + R/2$ (free effective RM) and 
prior RM$_{\rm fg} \sim \mathcal{U}(-3,3)\,\mathrm{rad\,m^{-2}}$ \citep{lofar23}. 

Results confirm 1T+2ED remains decisively preferred even with equal model flexibility: 
1Tk+2ED+RM$_{\rm fg}$ ($\Delta\ln Z = 5.87$; positive evidence against), 1T+1ED 
($\Delta\ln Z = 461.78$; very strong evidence against), and 1Tk+1ED+RM$_{\rm fg}$ 
($\Delta\ln Z = 462.30$; very strong evidence against). 

Thin leakage + 2ED is physically preferred (generally Bayes factors favour simpler model 
following Occam's Razor). Astrophysical components 
(RM $\approx 45$--$47\,\mathrm{rad\,m^{-2}}\gg3\,\mathrm{rad\,m^{-2}}$) remain unaffected 
by ionospheric modelling. We therefore retain the standard Sokoloff form \citep{Sokoloff98} for thick models, 
ensuring consistency with previous low-frequency studies.


Table~\ref{tab:best_fit_params} presents the best-fitting parameters of the models, where the rows are sorted by 2$\Delta$ln(Z) values.

\begin{table}[ht]
    \centering
    \caption{\textbf{Best-fit model parameters with 1$\sigma$ error bars. All  $p_{0,i}$ \text{ are intrinsic fractional polarizations (ranging from 0 to 1).}}}
    \label{tab:best_fit_params}
    \begin{tabular}{|l|l|l|l|l|l|}
    \hline
    \textbf{Model} &
    \parbox[t]{3cm}{\centering $p_{0,(1,2,3)}$ \\} &
    \parbox[t]{3cm}{\centering RM$_{(1,2,3)}$ \\ (rad m$^{-2}$)} &
    \parbox[t]{3cm}{\centering $\sigma_{\text{RM},(1,2,3)}$ \\ (rad m$^{-2}$)} &
    \parbox[t]{2cm}{\centering R \\ (rad m$^{-2}$)} &  
    \parbox[t]{3cm}{\centering $\psi_{0,(1,2,3)}$ \\ (deg)} \\
    \hline
    1T+2ED & $0.010^{+0.001}_{-0.001}$ & $-0.286^{+0.057}_{-0.054}$ & --- & --- & $126.395^{+14.848}_{-16.263}$ \\
    & $0.031^{+0.001}_{-0.001}$ & $-46.945^{+0.021}_{-0.021}$ & $0.010^{+0.011}_{-0.007}$ & --- & $75.817^{+6.066}_{-6.188}$ \\
    & $0.284^{+0.052}_{-0.045}$ & $-45.093^{+0.092}_{-0.087}$ & $0.274^{+0.010}_{-0.010}$ & --- & $111.247^{+19.024}_{-19.752}$ \\
    \hline
    1Tk+2ED & $0.013^{+0.002}_{-0.001}$ & --- & --- & $-0.235^{+0.043}_{-0.042}$ & $112.223^{+12.029}_{-12.115}$ \\
    & $0.031^{+0.001}_{-0.001}$ & $-46.943^{+0.020}_{-0.021}$ & $0.011^{+0.011}_{-0.008}$ & --- & $75.385^{+6.015}_{-6.077}$ \\
    & $0.284^{+0.053}_{-0.041}$ & $-45.087^{+0.092}_{-0.082}$ & $0.274^{+0.010}_{-0.010}$ & --- & $109.920^{+17.858}_{-20.157}$ \\
    \hline
    1T+1ED & $0.010^{+0.001}_{-0.001}$ & $-0.305^{+0.056}_{-0.058}$ & --- & --- & $132.074^{+16.169}_{-15.971}$ \\
    & $0.030^{+0.001}_{-0.001}$ & $-47.139^{+0.018}_{-0.018}$ & $0.012^{+0.013}_{-0.008}$ & --- & $134.033^{+4.886}_{-4.870}$ \\
    \hline
    1Tk+1ED & $0.012^{+0.001}_{-0.001}$ & --- & --- & $-0.241^{+0.045}_{-0.040}$ & $114.496^{+11.448}_{-12.719}$ \\
    & $0.030^{+0.001}_{-0.001}$ & $-47.138^{+0.017}_{-0.017}$ & $0.012^{+0.012}_{-0.008}$ & --- & $133.656^{+4.713}_{-4.781}$ \\
    \hline
    1ED & $0.030^{+0.001}_{-0.001}$ & $-47.137^{+0.017}_{-0.017}$ & $0.012^{+0.013}_{-0.008}$ & --- & $133.441^{+4.737}_{-4.745}$ \\
    \hline
    \end{tabular}
    \footnote{$p_{0}$ values are \textit{fractional} polarization; For example in model 1T+2ED, $p_{0,\text{thin}} \approx 0.01$ (1\% leakage), $p_{0,\text{ED1}} \approx 0.03$ (3\%), $p_{0,\text{ED2}} \approx 0.28$ (28\% of total I). The leakage component $\sim$ 0.01 (1\% of total I) is typical for LoTSS instrumental polarization \citep{lofar23}.}
\end{table}


The narrow astrophysical component ($p_{0,1} = 0.031 \pm 0.001$, 
RM$_1 = 46.94 \pm 0.02$ rad m$^{-2}$, $\sigma_{\rm RM,1} = 0.01 \pm 0.01$ rad m$^{-2}$) 
has negligible dispersion (consistent with $\sigma_{\rm RM,1} = 0$), resembling a 
foreground Faraday-thin screen. It could equivalently be modelled as thin, but allowing 
dispersion improves the fit near $\lambda^2 \approx 3{-}4$ m$^2$ without over-fitting 
(Bayes factor remains decisive).

The broad component ($p_{0,2} = 0.284 \pm 0.05$, RM$_2 = 45.09 \pm 0.09$ rad m$^{-2}$, 
$\sigma_{\rm RM,2} = 0.274 \pm 0.010$ rad m$^{-2}$) is separated in RM-space by 
$\Delta{\rm RM} \approx 2$ rad m$^{-2} \gg \sigma_{\rm RM,2}$, suggesting distinct 
magneto-ionic volumes (e.g., foreground cloud + lobe plasma). Its dispersion 
depolarizes it by $\exp(-2\sigma^2\lambda^4) \approx 21\%$ at $\lambda^2 = 3.2$ m$^2$ 
but $<0.5\%$ at $\lambda^2 = 6$ m$^2$, explaining why it dominates only high-frequency 
channels (hence absent in 1T+1ED fits).

The narrow component carries $\sim10\%$ of total $p_0$ but dominates mid-band signal, 
placing an upper limit $p_{0,{\rm broad}} \lesssim 0.3$ on undetected thicker emission 
($\sigma_{\rm RM} \gtrsim 10$ rad m$^{-2}$, invisible in LoTSS due to strong 
depolarization effects).

Beyond confirming that more than one external Faraday dispersion component is preferred, 
the fits yield several quantitative constraints that are directly relevant for future 
population studies (Table~\ref{tab:best_fit_params}). First, the two astrophysical 
components are tightly clustered in Faraday depth around RM $\sim 45{-}47$ rad m$^{-2}$ 
while requiring distinct dispersion widths $\sigma_{\rm RM}$ and intrinsic polarization 
fractions, indicating at least two magneto-ionic regions along the line of sight that 
are co-located in RM but differ in turbulence level or covering factor. Second, the 
relative amplitudes of these components imply that the more strongly dispersed screen 
contributes the majority of the depolarization across the LoTSS band, whereas the 
narrower component remains detectable even at the longest wavelengths, providing a 
concrete example of how LoTSS preferentially selects less-depolarized sub-components 
within otherwise complex sources.


\begin{figure}[htbp]
    \centering
    \includegraphics[width=1\textwidth]{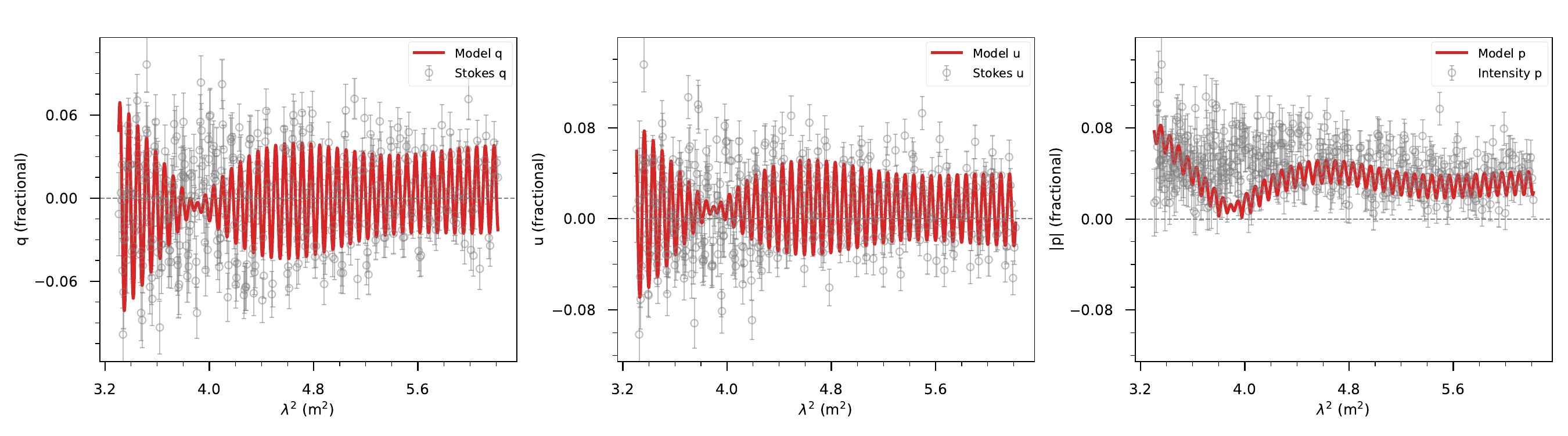}
    \caption{Fractional $q$, $u$, and polarized intensity $p$ plotted against $\lambda^2$ (top to bottom). Data (grey circles with error bars), best-fit 1T+2ED model (thick red line). The model effectively captures the observed depolarization trend across the LOFAR HBA band.}
    \label{fig:pqu}
\end{figure}

Figure~\ref{fig:pqu} shows the observed fractional Stokes parameters q, u, and the polarized fraction \(|p|\) as functions of 
wavelength squared. The over plotted curves correspond to the best-fit parameters of the \( \text{1T+2ED} \) model. 
Fractional Stokes parameters q, u generally exhibit sinusoidal or oscillatory variations due to Faraday rotation, but with decreasing amplitude as \( \lambda^2 \) increases. This is a hallmark of depolarization over wider wavelength ranges, polarization is partially cancelled due to the presence of multiple Faraday rotating regions or turbulent RM dispersion, causing the coherent addition of polarization vectors to diminish. We find the polarized fraction \( |p| = \sqrt{q^2 + u^2} \) versus \( \lambda^2 \) here show a hint of monotonic decline as \( \lambda^2 \) increases beyond a certain point. This is because increasing wavelength accentuates differences in RM along the line of sight or within the beam, causing more depolarization at longer wavelengths.


\begin{figure}[htbp]
    \centering
    \includegraphics[width=1\textwidth]{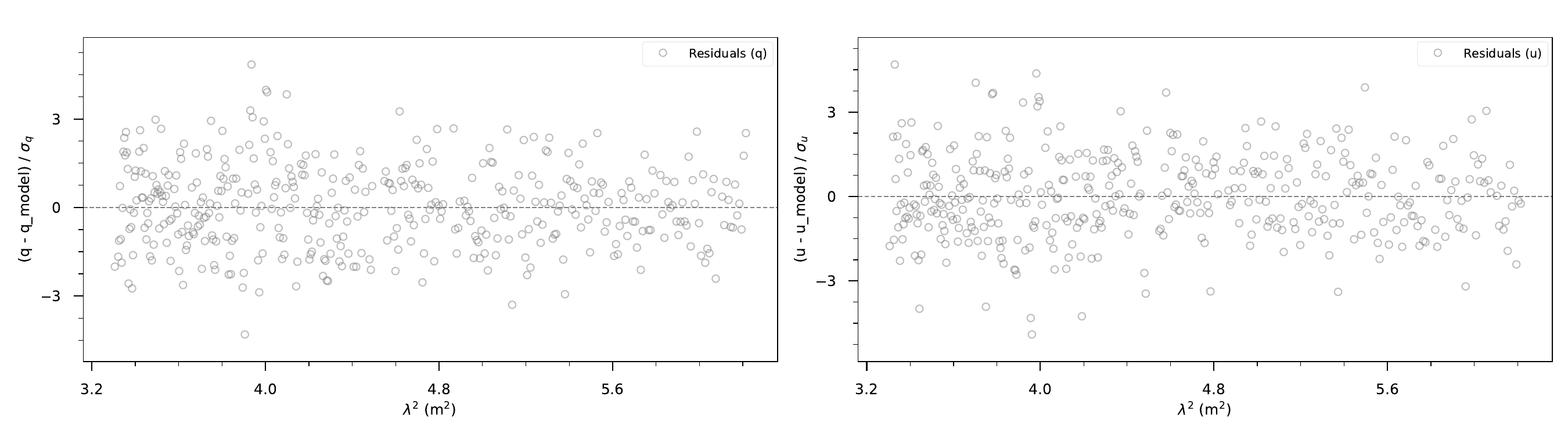}
    \caption{$q$, $u$ residuals normalized by measurement uncertainties versus $\lambda^2$.
    Residuals $\frac{Q_{\rm data}-Q_{\rm model}}{\sigma_Q}$ (blue) and 
    $\frac{U_{\rm data}-U_{\rm model}}{\sigma_U}$ (orange) show no strong 
    frequency-dependent correlations, consistent with underestimated noise 
    rather than unmodelled Faraday structure.}
    \label{fig:qures}
\end{figure}

To investigate the elevated $\chi^2_{\rm red}>1$, we examined $q$, $u$ residuals 
normalized by their measurement uncertainties, as a function of $\lambda^2$. Figure~\ref{fig:qures} shows noise-like residuals without strong frequency correlations, favouring underestimated noise over un-modelled Faraday structure. The discrepancy likely combines moderate model limitations with noise underestimation typical of interferometric imaging.


\begin{figure}[htbp]
    \centering
    \includegraphics[width=1\textwidth]{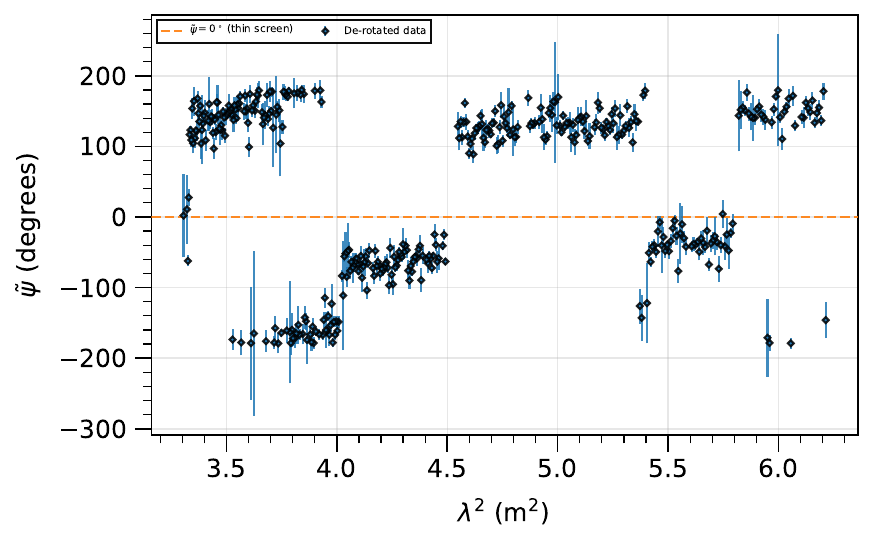} 
    \caption{De-rotated polarization angle $\tilde{\chi}(\lambda^2) = \chi(\lambda^2) + 47.15\,\lambda^2$ versus $\lambda^2$. The data exhibit extreme deviation from single thin-screen behavior ($\tilde{\chi} = 0^\circ$, orange line). A linear fit yields (not shown in the plot) residual RM = $+0.28 \pm 6.99$ rad m$^{-2}$, quadratic curvature ($\tilde{c} = 47.7 \pm 8.7$ deg m$^{-4}$) and reduced $\chi^2_\nu = 138.5$ ($>11\sigma$) against the linear model, providing strong evidence of complex Faraday structure, which requires advanced dispersion modelling.}
    \label{fig:derotpsi}
\end{figure}

Figure \ref{fig:derotpsi} presents the de-rotated polarization angle $\tilde{\psi}(\lambda^2) = \psi(\lambda^2) + 47.15\,\lambda^2$, where the LoTSS-DR2 catalogue RM$_{\rm cat}$ = -47.15 rad m$^{-2}$ has been used for de-rotation. Here, the raw polarization angles were first unwrapped and then de-rotated with $\pm$180$^\circ$ unwrapping. A single thin Faraday screen at this RM would produce a flat $\tilde{\psi}=0^\circ$ line. Instead, the data exhibit significant scatter and curvature around this reference.

Linear fitting yields a residual rotation measure RM$_{\rm res}$ = +0.28 $\pm$ 6.99 rad m$^{-2}$, statistically consistent with the catalogue value (consistent with dominant foreground screen removal). However, the quadratic term is highly significant ($>5\sigma$), indicating substantial departure from simple Faraday-thin screen behaviour.

The reduced $\chi^2_{\rm red} = 138.5$ further confirms systematic deviations from linear $\tilde{\psi} \propto \lambda^2$ behaviour, characteristic of either multiple discrete Faraday-thin screens along the line-of-sight, or Faraday dispersion within an extended magneto-ionic medium. This complexity also motivates our adoption of the 1T+2ED Faraday dispersion function modelling presented in Section \ref{sec:depolmodels}.


\begin{figure}[htbp]
    \centering
    \includegraphics[width=1\textwidth]{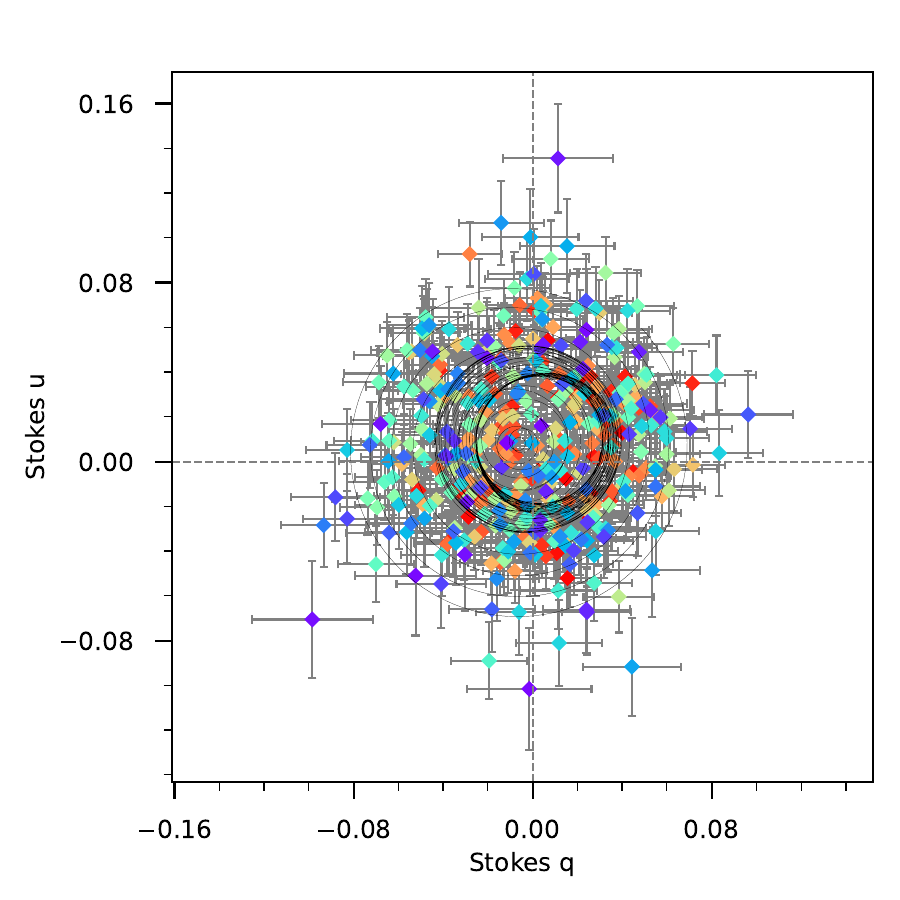}
    \caption{Polarization vectors in the complex plane (fractional Stokes parameters $q$ vs $u$) for ILTJ012215.21+254334.8, coloured by observing frequency (blue = high frequency/short $\lambda^2$, red = low frequency/long $\lambda^2$). Data points with $1\sigma$ error bars and the best-fit 1T+2ED model (solid line). The contraction of polarization vector radius from blue to red is direct evidence of wavelength-dependent depolarization. A Faraday-thin source would trace a circular arc at constant radius.}
    \label{fig:qvsu}
\end{figure}

Figure \ref{fig:qvsu} illustrates depolarization via vector contraction in the $q$-$u$ plane. High-frequency (blue) points cluster near the outer radius, indicating strong, coherent polarization. Low-frequency (red) points cluster near the centre, indicating severe depolarization. This trend is the same depolarization seen in Figure \ref{fig:pqu}, but represented geometrically rather than via polarized intensity vs $\lambda^2$.

\begin{figure}[htbp]
    \centering
    \includegraphics[width=1\textwidth]{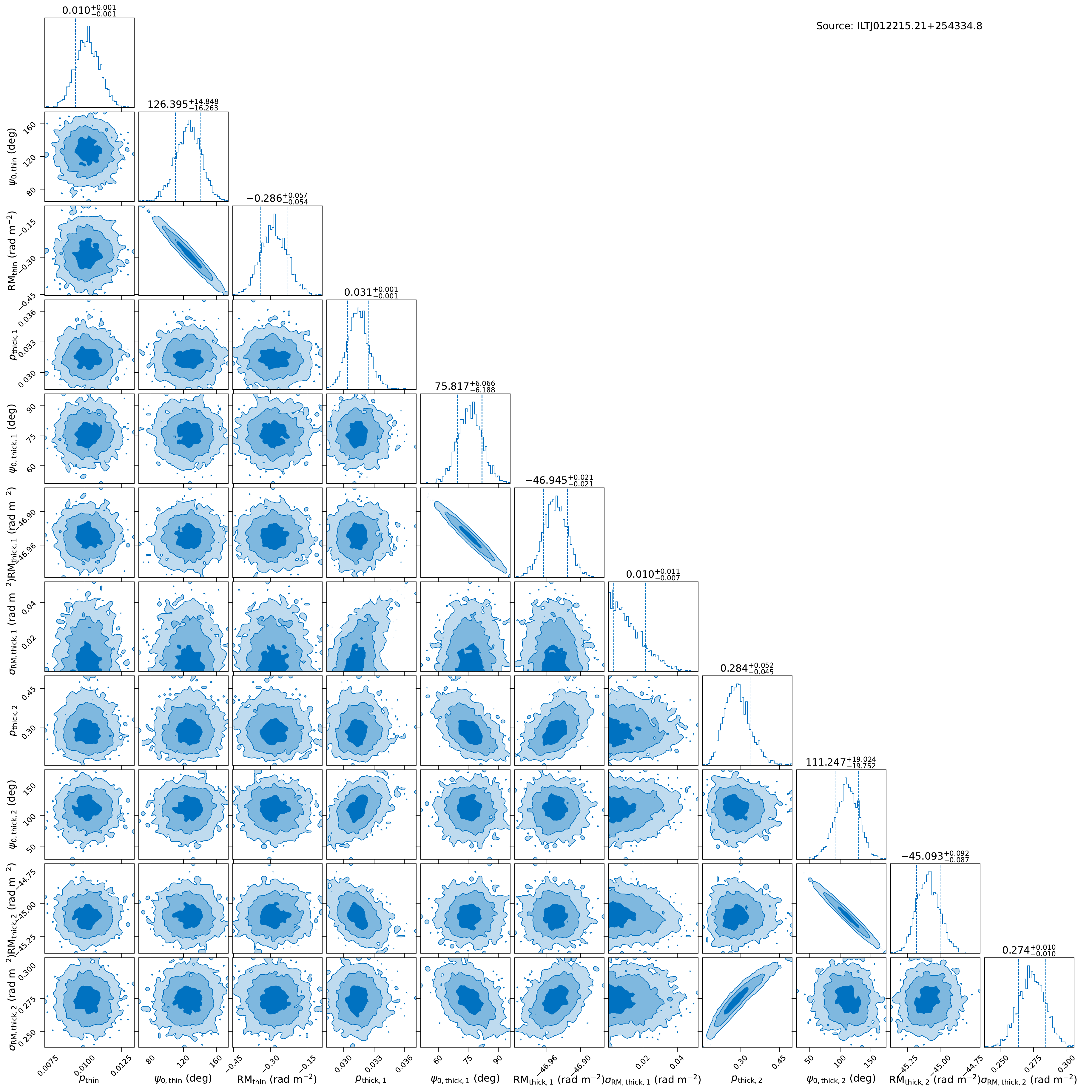}
    \caption{The posterior distribution of the best-fit model parameters are shown using one- and two-dimensional projections in a corner plot. This visualization illustrates both the individual distributions and the correlations among the parameters in the dataset. The 2D parameter plot shows the contours indicating 68\%, 95\% and 99.7\% confidence intervals. The diagonal sections display histograms for each parameter's distribution with $1\sigma$ (68\%) error bars. Off-diagonal panels display scatter plots that reveal correlations and covariances between pairs of parameters, illustrating how they jointly vary in the dataset.}
    \label{fig:mcmc}
\end{figure}

Finally, the posterior distribution of the best-fit model parameters along with $1-\sigma$ error bars is shown in Figure~\ref{fig:mcmc}. Note that the derived values of \( \psi_0 \) in Table~\ref{tab:best_fit_params} are not meaningful due to the lack of absolute calibration for the polarization angles.

\section{Summary and Conclusion}
\label{sec:discuss}
In this study, we have investigated the depolarization properties of the radio galaxy ILTJ012215.21-254334.8 using polarization data from the LOFAR Two-metre Sky Survey (LoTSS) DR2 catalogue. An important outcome of this exercise is that, even using LoTSS HBA data alone, the Bayesian qu-fitting framework can discriminate between one-, two-, and three-component depolarization models and deliver well-defined posteriors for the dominant external dispersion components. This shows that LoTSS-DR2 survey data alone can robustly constrain multi-component Faraday depolarization models for typical sources in the DR2 catalog.

To model the observed polarization behaviour, we employed a suite of depolarization models - including Faraday-thin, Faraday-thick, and external Faraday dispersion components - and combined them in a Bayesian qu-fitting framework. Statistical comparison using Bayesian evidence and reduced chi-squared metrics identified the three-component \texttt{1T+2ED} model as providing the best fit to the data. This model consists of a Faraday-thin component interpreted as instrumental polarization leakage (near zero RM) and two external Faraday dispersion components centred at $\mathrm{RM} \approx -47\,\mathrm{rad} \ \mathrm{m}^{-2}$, consistent with values derived from the LoTSS DR2 catalogue. A relatively higher reduced chi-squared (\(\chi^2_{\rm red}\)) value of 2.12 suggests that the errors have been underestimated, or that there are systematic discrepancies unaccounted for by the model. Pinpointing the exact cause of the difference is difficult because it is probably due to both the model not capturing all details and the noise or errors being underestimated. The discrepancy likely arises from a combination of these two factors.

The necessity for multiple external depolarization components suggests a complex magneto-ionic medium with significant turbulence and inhomogeneities. Direct observational evidence for depolarization is the monotonic decline in $p(\lambda^2)$ shown in Figure \ref{fig:pqu}, which represents wavelength-dependent loss of coherence in the polarized signal. Posterior analysis further supports the role of both turbulence and line-of-sight structure in shaping the observed polarization. Selecting a source with a relatively large projected linear size (335 kpc) has enabled a robust investigation of depolarization effects in a typical low-frequency radio galaxy. The polarized intensity, RM properties measured are consistent with the LoTSS DR2 sample.

An important limitation of the present analysis is its intrinsic insensitivity to very Faraday-thick components. Because the LoTSS HBA band is confined to low frequencies, any emission with \(\sigma_{\mathrm{RM}}\) significantly larger than a few rad m\(^{-2}\) would be almost completely depolarized across the band and would not contribute to the observed \(q,u\) spectra, even if intrinsically bright and highly polarized. The fitted external dispersion components therefore probe only the \textit{moderately} Faraday-thin or thick part of the magneto-ionic environment; additional, much more heavily dispersed screens may be present but remain invisible to LoTSS and can only be constrained by combining these data with higher-frequency polarization measurements.

In conclusion, our results demonstrate the efficacy of low-frequency polarization observations, combined with quantitative qu-fitting, in separating instrumental effects from astrophysical depolarization and Faraday rotation signatures. The best-fitting model reveals the importance of multi-component depolarization in a typical radio galaxies of the LoTSS DR2 catalogue, reflecting the inherent complexity of their magnetized environments, usually found in dense cluster environments. In future, extending this analysis to the full FR-I sample in the LoTSS DR2 catalogue, will help establish broader statistical trends and further constrain the interplay between turbulence, environmental density, source size, and polarization properties in extragalactic radio sources. Thus, while this single-source study does not aim to provide a complete physical model of ILTJ012215.21-254334.8, it serves as a proof of concept for extracting robust Faraday-depolarization constraints from LoTSS-only data, and it quantifies both the power and the limitations of such an approach. The derived multi-component structure in Faraday depth, the required level of RM dispersion, and the model-selection behaviour provide concrete benchmarks for a forthcoming systematic analysis of the full FR-I subsample in the LoTSS-DR2 RM Grid.

\section*{Acknowledgments}
SSP would like to thank Banwarilal Bhalotia College, Asansol, under Kazi Nazrul University, for facilitating the M.Sc. Physics program. This research forms part of the M.Sc. thesis work conducted in the final semester. The authors appreciate ChatGPT for helping with AI-driven copy editing and improving the manuscript's clarity. AG acknowledges the financial support received through the SERB-SURE grant (SUR/2022/000595) from the Science and Engineering Research Board (SERB), a statutory body under the Department of Science and Technology (DST), Government of India. AG would also like to thank IUCAA, Pune, for their support through the Associateship Programme and for providing access to their computational facilities. We would like to sincerely thank the reviewers for their thoughtful and constructive feedback on our manuscript.


\bibliographystyle{apsrev4-1}
\bibliography{main}

\end{document}